\documentclass[twocolumn,showpacs,preprintnumbers,amsmath,amssymb]{revtex4}
\usepackage{graphicx}% Include figure files
\usepackage{dcolumn}% Align table columns on decimal point
\usepackage{bm}% bold math

\begin{document}

\preprint{APS/123-QED}

\title{Design and optimization of one-dimensional ferrite-film based magnonic crystals}

\author{A.V.~Chumak}

 \email{chumak@physik.uni-kl.de}
 \altaffiliation[\\ Also at ]{National Taras Shevchenko University of Kiev, Ukraine.}

\author{A.A.~Serga}

\author{S.~Wolff}

\author{B.~Hillebrands}

\affiliation{Fachbereich Physik, Nano+Bio Center, and
Forschungszentrum OPTIMAS, Technische Universit\"at Kaiserslautern,
67663 Kaiserslautern, Germany}

\author{M.P.~Kostylev}
\affiliation{School of Physics, University of Western Australia, Crawley, Western Australia 6009,
Australia}

\date{\today}

\begin{abstract}

One-dimensional magnonic crystals have been implemented as gratings
of shallow grooves chemically etched into the surface of
yttrium-iron garnet films. Scattering of backward volume
magnetostatic spin waves from such structures is investigated
experimentally and theoretically. Well-defined rejection frequency
bands are observed in transmission characteristics of the magnonic
crystals. The loss inserted by the gratings and the rejections bands
bandwidths are studied as a function of the film thickness, the
groove depth, the number of grooves, and  the groove width.  The
experimental data are well described by a theoretical model based on
the analogy of a spin-wave film-waveguide with a microwave
transmission line. Our study shows that magnonic crystals with
required operational characteristics can be engineered by adjusting
these geometrical parameters.

\end{abstract}

% PACS, the Physics and Astronomy Classification Scheme.

\maketitle

\section{introduction}

Periodically structured magnetic materials such as magnonic crystals
(MC) attract special attention in view of their
applicability for both fundamental research on linear and nonlinear
wave dynamics in artificial media, and for signal processing in
microwave frequency range \cite{MC review}. Similar to sound and
light in sonic and photonic crystals, the dispersion characteristics
of spin waves in magnonic crystals are strongly modified with
respect to uniform media. This results in the appearance of frequency band
gaps \cite{Nikitov, BH Kolodin} wherein spin-wave propagation is forbidden.
A large variety of nonlinear spin-wave phenomena, as well as
the dependency of the spin-wave properties both on the magnitude and
orientation of a bias magnetic field determine the wide tunability
of operational characteristics of the magnonic crystals and their
potential for design of microwave filters, switchers, current
controlled delay lines, power limiters, etc \cite{MC review}.

Depending on the required insertion loss, operating frequency range,
dimensions, temperature stability, and other performance
specifications, periodic structures can be fabricated
from either ferrite or ferromagnetic substances by means
of geometric structuring \cite{MC review, MC skyes, MC MSSW,
MC FVMSW, Gubbiotti1, Gubbiotti2, Ogrin, blade}, metal deposition
\cite{metalization}, ion implantation \cite{ionic implantation},
local variations of the bias magnetic field \cite{field modulation},
to name but a few.

At the present time, the smallest out-of-band insertion loss in
conjunction with the deepest rejection bands have been observed in
the experiments with geometrically structured yttrium-iron-garnet
(YIG) single crystal ferrite films grown on a gallium-gadolinium
substrate by means of liquid-phase epitaxy. In this unique material
\cite{Saga} the lifetime of spin-wave excitations can exceed a
couple of hundreds nanoseconds, and the spin-wave propagation path
reaches a few centimeters.

The surface magnetostatic waves \cite{Damon-Eshbach} (wave
propagation direction is perpendicular to the magnetic field applied
in the film plane) and the forward magnetostatic waves
\cite{Damon-VanDeVaart} (external field is oriented perpendicular to
the film plane) were previously studied theoretically and
experimentally in such periodical structures \cite{MC review, MC
skyes, MC MSSW, MC FVMSW}.

In our recent paper \cite{APL_MC} we presented the first
experimental and theoretical results on scattering of backward
volume magnetostatic waves (wave propagation direction is parallel
to the magnetic field applied in the film plane)
\cite{Damon-Eshbach} from an one-dimensional structure with periodic
changes of the YIG film thickness.

In the present paper we present a detailed study of this kind of
magnonic crystal. Its main characteristics, such as insertion loss
in the rejection bands, parasitic loss in the transmission bands,
and the frequency bandwidth of the rejection bands, were
investigated for crystals having different groove depths, widths,
and groove numbers. Our results show these parameters can be used to
optimize the design of magnonic crystals.

\section{technology and experimental measurements}

To fabricate magnonic crystals, 5.5\,$\mu$m and 14\,$\mu$m-thick YIG
films,  which were epitaxially grown in the (111) crystallographic
plane, were used. Photolithographic patterning followed by hot
orthophosphoric acid etching was used to form the grooves. The
lithography was based on a standard photoresist AZ~5214E hardened by
UV irradiation and baking which makes it stable against hot 160
$\mathrm{^o}$C orthophosphoric acid. Using this procedure we
patterned arrays of $N$ parallel lines ($N = 10$ and $N = 20$) of
widths $w = 30$\,$\mu$m and interline spacings of 270\,$\mu$m.
Alternatively, arrays where $w = 10$\,$\mu$m spaced by 290\,$\mu$m
were also prepared. In all cases, the lattice constant was $a =
300\,\mu$m. The grooves were transversely oriented with respect to
the spin-wave propagation direction. In order to study the
dependence of crystal characteristics on the groove depth $\delta$
the grooves were etched in 100\,nm steps from 100\,nm to 2\,$\mu$m.
The groove depth was controlled by the etching time (etch rate was
$\sim 1\, \mu $m/min) and measured using a profilometer. Anisotropic
chemical etching of the YIG crystal structure by orthophosphoric
acid \cite{etching} was observed: the etch rate parallel to the film
plane was approximately ten times larger than in the perpendicular
direction, so the final groove depth profile along the direction of
wave propagation had a trapezoidal shape.

\begin{figure}
\includegraphics[width=0.9\columnwidth]{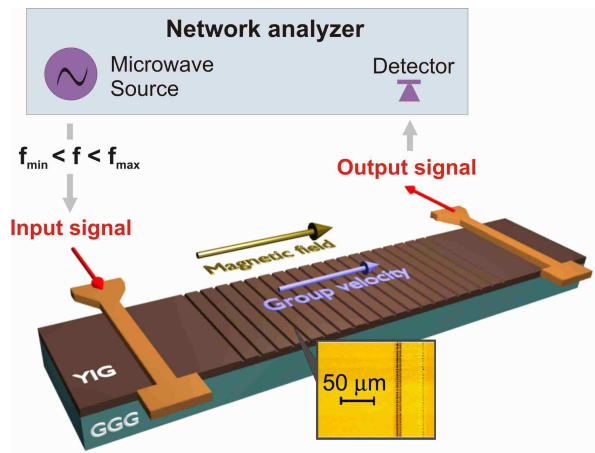}
\caption{\label{setup} (Color online) Scheme of a magnonic crystal
structure and of the measurement setup used in the experiments.}
\end{figure}

The inset in Fig.~\ref{setup} shows the microscope image of the
grooves structure of the magnonic crystal with the groove depth
$\delta = 500$\, nm and groove width $w = 30$\,$\mu$m. Anisotropic
etching can be clearly seen.

In order to excite and receive the dipolar spin waves, two
microstrip antennas were placed 8\,mm apart, one in front of the
grating, and one behind it (see Fig.~\ref{setup}). A bias magnetic
field of 1845\,Oe was applied in the plane of the YIG film stripe,
along its length and parallel to the direction of spin-wave
propagation. Under these conditions, backward volume magnetostatic
wave (BVMSW) propagation occurs. A continuous-wave microwave signal
was applied from a network analyzer (see Fig.~\ref{setup}) to the
input antenna, and BVMSW transmission characteristics were measured.
The microwave signal power was maintained at 1\,mW in order to avoid
any non-linear processes.

\section{Theoretical model}

In a general case, a theoretical description of the scattering of
dipole spin waves from inhomogeneities is given by a singular
integral equation
\begin{eqnarray}\label{grin}
{\bf m}({\bf r})  =  4 \pi \hat{\kappa}(f,{\bf H}_{\it i}({\bf
r}),{\bf M}({\bf r})) \cdot \nonumber\\
 \left( \int_{V}\hat{\rm G}({\bf
r}-{\bf r}') {\bf m}({\bf r'})d{\bf r}'+{\bf h}({\bf r})\right),
\end{eqnarray}
where $\bf{m}$ is the dynamic magnetization, ${\bf h}$ is the
microwave field of the input antenna, $\hat{\rm G}({\bf r}-{\bf
r}')$ is the Green's function of dipole magnetic field, and
$\hat{\kappa}(f,{\bf H}_{{\it i}}({\bf r}),{\bf M}({\bf r}))$ is the
tensor of the microwave magnetic susceptibility which depends on the
spin-wave carrier frequency $f$ and on the magnetic parameters of
the film. The latter are functions of the position ${\bf r}$, since
the grooves induce inhomogeneity in the internal static field ${\bf
H}_{\it i}$ and in the equilibrium magnetization ${\bf M}$. The
integration is carried over the volume of the magnetic structure,
thus the inhomogeneity of the film thickness is taken into account.
A simple approximate solution to Eq.(1) for the reflected wave can
be obtained  in the frame of the first Born approximation (see
\cite{kostylev} for details of the approach). Neglecting magnetic
damping in the medium far in front of the groove lattice $x<<x'$ for
the complex reflection coefficient $R$, one obtains
\begin{equation}
R(k)=i\frac{w}{d_0} F(k,d_0,\delta)
\frac{\sin{(kw)}}{kw}\sum_{n=0}^N \exp (-2ikna),
\end{equation}
 where $F(k,d_0,\delta)=\frac{[\exp(-kd_0)(\exp(k\delta)-1)-(\exp(-k\delta)-1)]}{kd_0}\approx
2\delta/d_0-k\delta$ for $kd_0<<1$, where $k$ is the wavenumber of
the incident spin wave, $d_0$ is the unpatterned film thickness,
$\delta$ is the groove depth, $w$ is the groove width, $a$ is the
lattice period, and $N$ is the total number of grooves in the groove
lattice. This formula is in agrement with Bragg's diffraction law.
The maxima of reflection occur at $k=n\pi/a$ or for the wavelengths
$2a/n$. With increasing $N$ the depth of the reflection bands
increase and their wavenumber widths decrease. Furthermore, the
reflection grows with an increase in $w$ and $\delta/d_0$, which is
consistent. However, according to this formula, the reflection
should decrease with increasing $k$, which is in contradiction with
the experimental results presented here (see Section IV). Thus a
more detailed model is necessary.

To build a more appropriate model,  we first notice that the
quasi-1D dipole field \cite{guslienko} of the lowest BVMSW thickness
mode decays within a distance of a few film thicknesses from its
source. Because the width of the grooves $w$ is much smaller than
$a$, the spin wave travels as an eigenmode of a continuous film of
thickness $d_0$ through most of the lattice (\emph{i.e.} between
grooves). For the sections where the thickness is $d_0$ the integral
equation reduces to a simple formula (see Eq.~(50) in
\cite{kostylev}) which shows that between the grooves, the
transmitted and reflected waves only accumulate phase and decay (due
to intrinsic magnetic damping). Thus, in order to describe the
formation of stop bands, one has to consider the scattering of a
BVMSW from just one groove. The effect of multiple consecutive
grooves is obtained by cascading the structure period using matrices
of scattering T-parameters and taking interference effects into
account.

The most direct way to proceed is to solve this two-dimenstional
singular integral equation numerically to obtain the scattering
characteristics of a BVMSW scattering from a single groove. However
this work is beyond the scope of this paper. Another way to treat
scattering from a single groove was previously suggested in Refs.
\cite{kalinikos, Maeda}. In our short paper \cite{APL_MC}, we
adapted this method to the case of a BVMSW. For completeness, we
reproduce our theory here, but in more detail. We first consider the
grating as a periodical sequence of sections of regular transmission
lines with different propagation constants (different spin-wave
wavenumbers) for the same carrier frequency. We neglect the fact
that the groove edges are oblique, and consider the groove
cross-section as a rectangle with the same depth and having the same
area. In order to describe spin wave transmission through the
magnonic crystal we use T-matrices which describe the relation
between the amplitudes of the wave incident onto an inhomogeneity
and reflected from it \cite{transm line}. The T-matrix
$\mathrm{T^{(1)}}$ for a section of unstructured film (section of
film between neighboring grooves) of length $a-w$ has diagonal
components only:
\begin{equation}\label{t1}
\mathrm{T^{(1)}} =
\begin{pmatrix} e^{(-ik+k''_0)
(a-w)} & 0 \\ 0 & e^{(ik-k''_0) (a-w)} \end{pmatrix},
\end{equation}
where $k$ is the spin-wave wavenumber in the unstructured film,
$k''_0 = \gamma \Delta H /(2 v_\mathrm{gr})$ is the rate of the
spin-wave spatial damping, $\gamma$ is the gyromagnetic ratio,
$\Delta H$ is the ferromagnetic resonance linewidth, and
$v_\mathrm{gr}$ is the spin-wave group velocity.

Similarly, the T-matrix $\mathrm{T^{(3)}}$ for a regular spin-wave
film waveguide with a thickness $d=d_0-\delta$ is
\begin{equation}\label{t3}
\mathrm{T^{(3)}} =
\begin{pmatrix} e^{(-ik +
k''_\mathrm{g})w d_\mathrm{0} / d}  & 0 \\ 0 & e^{(+ik -
k''_\mathrm{g})w d_\mathrm{0} / d} \end{pmatrix},
\end{equation}
where $k''_\mathrm{g}$ is the spin-wave damping rate for the groove.
Here we use the fact that the BVMSW dispersion law for small
wavenumbers $kd \ll 1$ is practically linear. Therefore, the
spin-wave wavenumber in the grooves is $k d_\mathrm{0}/d$. To
describe the loss increase in the pass bands with increasing groove
depth, we introduce an empirical parameter $\zeta$ which accounts
for larger contribution of two-magnon scattering processes in the
areas which underwent anisotropic etching \cite{etching}. Then the
damping rate in the grooves can be expressed as $k''_\mathrm{g} =
k''_0 (1+\zeta \delta/d_0)$.

At the edges of the grooves the  incident wave is partially
reflected back. This is accounted for by the T-matrices for the
groove edges. The matrix for the front edge is $\mathrm{T^{(2)}}$,
and that for the rear edge is $\mathrm{T^{(4)}}$. Following
\cite{transm line}, the transmission coefficient through the
junction is $1-\Gamma$. Then one obtains:
\begin{equation}\label{t2}
\mathrm{T^{(2)}} =
\begin{pmatrix} (1-\Gamma)^\mathrm{-1}  & \Gamma (1-\Gamma)^\mathrm{-1} \\
\Gamma (1-\Gamma)^\mathrm{-1} & (1-\Gamma)^\mathrm{-1}
\end{pmatrix},
\end{equation}
\begin{equation}\label{t4}
 \mathrm{T^{(4)}} =
\begin{pmatrix} (1+\Gamma)^\mathrm{-1}  & -\Gamma (1+\Gamma)^\mathrm{-1}\\
-\Gamma (1+\Gamma)^\mathrm{-1} & (1+\Gamma)^\mathrm{-1}
\end{pmatrix}.
\end{equation}
Using the property of T-matrix multiplication one finds a T-matrix for one period
of the structure:
\begin{equation}\label{t_total}
 \mathrm{T} = [\mathrm{T^{(1)}} \cdot
\mathrm{T^{(2)}}\cdot \mathrm{T^{(3)}} \cdot \mathrm{T^{(4)}}].
\end{equation}
To obtain the T-matrix for a magnonic crystal with $N$ grooves
$\mathrm{T^\mathrm{mc}}$, one has to raise $\mathrm{T}$ to the
${N}$-th power:
\begin{equation}\label{t_mc}
 \mathrm{T^\mathrm{mc}} = [\mathrm{T^{(1)}} \cdot
\mathrm{T^{(2)}}\cdot \mathrm{T^{(3)}} \cdot \mathrm{T^{(4)}}]^N.
\end{equation}
The most important operational parameter of magnonic crystals is the
power transmission coefficient. It can be determined as
$P_\mathrm{tr} = 1/\mid T^\mathrm{mc}_{11} \mid ^2 = 1/\mid
T^\mathrm{mc}_{22}\mid ^2$, where $T^\mathrm{mc}_{11}$ and
$T^\mathrm{mc}_{22}$ are the matrix elements.

In order to make use of this theory, one has to specify the form of
the reflection coefficient $\Gamma$. The model we suggest is based
on the analogy of the change in the film waveguiding properties to a
change in the characteristic impedance $Z$ of a microwave
transmission line \cite{kalinikos}. The expression for the complex
reflection coefficient for a junction of two microwave lines with
characteristic impedances $Z_0$ and $Z$ can be written as
\cite{transm line}:
\begin{equation}\label{G_transLine}
 \Gamma^\mathrm{tr. line} = \frac{Z - Z_0}{Z + Z_0}.
\end{equation}
We assume that the change of the characteristic impedance of a
spin-wave waveguide arising from the change of YIG-film thickness is
due to a change of the film's effective inductance. Then the
characteristic impedance is linearly proportional to the propagation
constant (to the spin-wave wavenumber in our case), and we obtain a
formula for the reflection coefficient for a wave incident onto the
edge of a groove from the unstructured section of the film:
\begin{equation}\label{G_transLine}
\Gamma = \eta \frac{d_0 - (d_0 - \delta)}{d_0 + (d_0 - \delta)}=
\eta \frac{\delta}{2d_0-\delta},
\end{equation}
where $\eta > 1$ is a phenomenological parameter introduced in this
formula to account for eventual factors which were not taken into
account in this simplistic model.

For the wave incident onto the same junction in the reverse
direction, $\Gamma_{-}=-\Gamma$, which has already been taken into
account in the expressions for the T-matrices above. The expression
$\Gamma_{-}=-\Gamma$ means that the wave phase change due to
reflection from one and another edge of groove is $\pi$.

\section{Results and discussion}

\begin{figure}
\includegraphics[width=0.95\columnwidth]{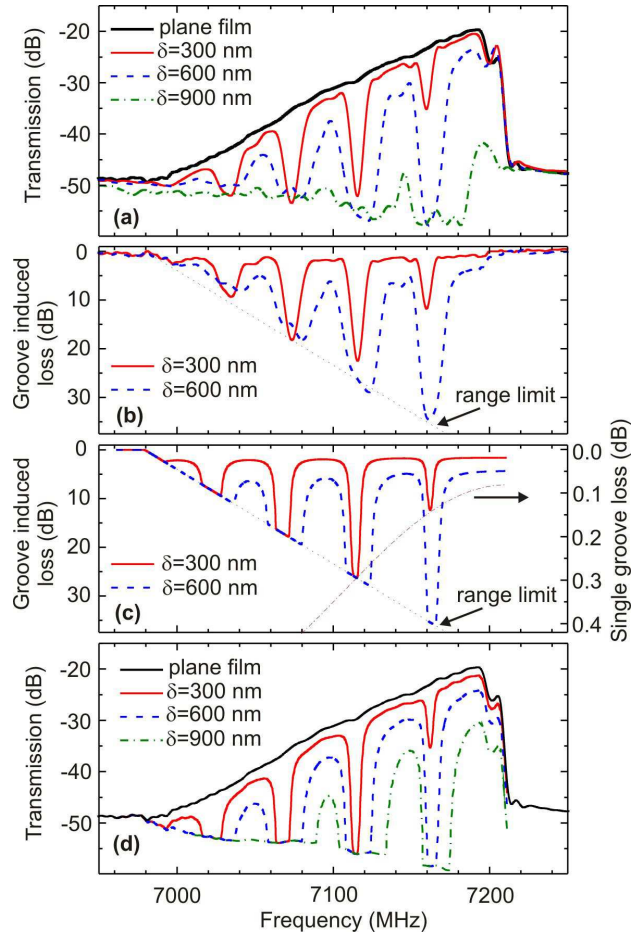}
\caption{\label{main_res} (Color online) (a) - BVMSW microwave
transmission characteristics for an unstructured film (bold line)
and for magnonic crystals with different groove depths $\delta$; (b)
- measured transmission loss inserted by the magnonic crystal
structure; (c) - calculated loss; (d) - calculated transmission
characteristics. Parameters of calculation: groove number $N = 20$,
width of grooves at their bottom $w = 30$\,$\mu$m, lattice constant
$a = 300\,\mu$m, film thickness $d_\mathrm{0} = 5.5$\,$\mu$m,
saturation magnetization $4 \pi M_\mathrm{0} = 1750$\,G, bias
magnetic field $H_\mathrm{0} = 1845$\,Oe, efficiency coefficient
$\eta=6$, resonance line width $\Delta H = 0.5$\,Oe, surface damage
coefficient $\zeta = 30$. In (c), the dash-dotted line shows the calculated
loss inserted by a single groove of 300\,nm in depth.}
\end{figure}

The experimental BVMSW transmission characteristics for the
unstructured film and for the magnonic crystals with $\delta$~=~300,
600 and 900\,nm measured with a network analyzer are shown in
Fig.~\ref{main_res}(a). The initial YIG film thickness $d_0$ is
5.5~$\mu$m. The results are shown for the groove number $N = 20$ and
the groove width $w = 30$\,$\mu$m. The BVMSW transmission
characteristic for the unstructured film has a maximum just below
the point of ferromagnetic resonance. One sees that the insertion
loss is $\approx 20$~dB and is determined by the energy
transformation efficiency by the input and the output antennas and
by the spatial decay of spin waves during their propagation in the
space between the antennas. As BVMSW frequency band is bounded above
by the ferromagnetic resonance frequency no spin-wave propagation
occurs for higher frequencies. With decreasing frequency the BVMSW
excitation and reception efficiencies drop. This drop is due to the
finite width of the antennas in the direction of BVMSW propagation.
Thus for very low and very high frequencies, no spin waves can be
excited and the insertion loss ($\approx 50$~dB) is determined by
the direct electromagnetic leakage from the input to the output
antenna. A small separate peak at the right-hand edge of the
transmission characteristic is due to excitation of the first width
standing mode of film \cite{width mode}.

In order to remove the influence of the antennas and of the
spin-wave spatial decay we calculate the difference between the
logarithms of the transmission characteristics for the unstructured
and the structured films. Figure~\ref{main_res}(b) shows the
obtained dependence. The additional dotted straight line in this
figure indicates the limit of the dynamic range of our experimental
setup. The latter is found as the difference in the transmission
characteristics of the microstrip antenna structure that is covered
or uncovered by a continuous YIG film.

From Fig.~\ref{main_res}(b) one sees that a lattice of 20 grooves as
deep as 300\,nm leads to the appearance of a set of rejection bands
(or transmission gaps), where spin-wave transmission is highly
reduced. According to the condition for Bragg reflection,
higher-order rejection bands correspond to larger spin-wave
wavenumbers. In the case of BVMSWs, the latter corresponds to lower
frequencies. From the depths and the frequency widths $\Delta{f}$ of
the gaps, one sees that the efficiency of the rejection increases
with increasing order of Bragg reflection. This suggests that BVMSW
with smaller wavelengths are more sensitive to the introduced
inhomogeneities.

Both Fig.~\ref{main_res}(a) and ~\ref{main_res}(b) demonstrate that
an increase in $\delta$ leads to an increase in the rejection
efficiency and in the frequency width of rejection bands
$\Delta{f}$. Additionally, a small frequency shift of the minima of
transmission towards higher frequencies is observed, as well as an
increase in insertion losses in the transmission (i.e. allowed)
bands. For $\delta = 900$\,nm the insertion loss in the whole
spin-wave band is so important that almost no spin-wave propagation
is observed (see Fig.~\ref{main_res}(a)) for the film of 5.5~$\mu$m
thickness.

The results of our numerical computation of $\mathrm{T^\mathrm{mc}}$
 are shown in Fig.~\ref{main_res}(c). One sees that
this model provides qualitative  agreement with all the
experimentally observed trends. In particular, it correctly predict
the observed increase in the rejection efficiency with increasing
$k$. The calculated efficiency of the rejection also increases with
increasing rejection band order. In order to prove this we calculate
the transmission characteristic for a structure which consists of
only one 30~$\mu$m wide and 300\,nm deep groove (see the dash-dotted
line in Fig.~~\ref{main_res}(c)). One sees that the efficiency of
spin-wave reflection from one groove increases with increase in spin
wave wavelength. It is worth noting that the transmission loss
inserted by one groove is about 0.1\,dB. Thus, almost all the energy
of the spin waves is transferred through the groove. Only a very
small part (about 3 \%) is reflected back.

Fig.~\ref{main_res}(c) shows that our computation gives the correct
shape of the transmission characteristics. In particular, there is a
good agreement with the frequency widths $\Delta{f}$ of rejection
bands and in their frequency shifts upwards with increasing
$\delta$. Furthermore, the model properly describes the increase in
the parasitic insertion loss in the transmission bands with
increasing groove depth.

The calculated influence of the groove array on the transmission
characteristic of the magnonic crystal is shown in
Fig.~\ref{main_res}(d). The presented curves were found as a
difference of the experimental transmission characteristic for a
plane film (bold line in Fig.~\ref{main_res}(a)) and the calculated
groove induced loss (Fig.~\ref{main_res}(c)). As a result,
Fig.~\ref{main_res}(d) are a theoretical analogue to
Fig.~\ref{main_res}(a), where experimental results are presented.
From comparison of the figures one can concludes that proposed
theoretical model describes experimental results especially good for
the small values of the groove depth. With increasing of the groove
depth the disagreement increases because of the theoretical model
limitations.

\subsection{Influence of the groove depth}

\begin{figure}
\includegraphics[width=0.9\columnwidth]{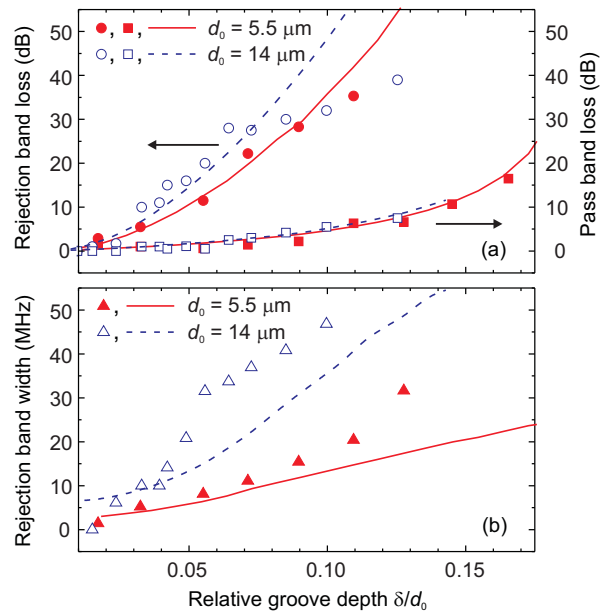}
\caption{\label{diff_d} (color online) (a) - insertion loss in
the first-order rejection band (circles) and the pass band (squares) as a function of
the relative groove depth $\delta/d_0$; lines show corresponding
calculated dependencies. (b) - experimental (triangles) and
theoretical (lines) frequency width of the first-order rejection band as a function of
the relative grooves depth. In both panels, filled red symbols and solid red lines
show dependencies for the film $d_0 = 5.5\,\mu$m thick;
opened blue symbols and dashed blue lines are
for $d_0 = 14\,\mu$m. Number of grooves
$N = 20$; lattice constant $a = 300\,\mu$m; groove width $w =
30$\,$\mu$m.}
\end{figure}

The groove depth has a profound influence on the transmission
characteristics of the fabricated magnonic crystals (see
Fig.~\ref{main_res}). In order to investigate this effect in
Fig.~\ref{diff_d} we plot the insertion loss for the first-order
rejection band, its frequency width, and the parasitic loss in the
first transmission band for magnonic crystals of different
thicknesses $d_0$ and with different groove depths. The central
frequency of the first-order gap is 7160\,MHz. From
Fig.~\ref{diff_d}~(a) one sees that the rejection efficiency
strongly increases with increasing the relative groove depth
$\delta/d_0$. An increase in the parasitic loss in the transmission
bands is also observed. By comparing the losses in the rejection and
transmission bands, we can estimate the optimal value of
$\delta/d_0$. We define this optimum value as a situation in which
rejection is efficient but parasitic loss is still small (around
3~dB). For both films we find an optimal value of $\delta/d_0
\approx 0.1$.

Both our calculation and measurements for different film thickness
show that parasitic losses in transmission bands are determined by
the relative groove depth only. The situation with the rejection
efficiency is more complicated. From Fig.~\ref{diff_d}~(a) it can be
seen that the experimental dependencies are slightly different for
different film thicknesses. For the same relative groove depth,
rejection is larger  for thicker films. However, with increasing
$\delta$ this difference between films with different thicknesses
diminishes, and for the largest values of $\delta/d_0$ the
experimental dependencies for different thicknesses collapse. This
suggests that the relative groove depth is the leading parameter for
the magnonic crystal optimization.

It is worth noting that the behavior of the calculated curves is
slightly different. This suggests that the accuracy of our simple
theory decreases with increasing groove depth. Processes not taken
into account in our model probably become more prominent with larger
$\delta/d_0$.

Fig.~\ref{diff_d}~(b) shows the experimental and calculated
frequency width of the first rejection band. The frequency width was
measured at the distance between the points for which the
transmitted signal intensity is halved. Obviously, the same Bragg
condition for the spin-wave number in the maximum of rejection is
fulfilled for the magnonic crystals based on the films with
thickness 5.5\,$\mu$m and 14\,$\mu$m, because the array has the same
lattice constant 300\,$\mu$m in both cases. However, the spin wave
dispersion strongly depends on film thickness \cite{Damon-Eshbach}.
For small wave wavenumbers its slope increases with increase in
$d_0$. This results in the wider rejection bands observed for the
thicker magnonic crystal.

From Fig.~\ref{diff_d}~(b) one sees that the width of the rejection
band for the optimal relative groove depth 0.1 is 15~MHz and 45~MHz
for the 5.5\,$\mu$m and 14\,$\mu$m thick films, respectively. From
the point of view of optimizing magnonic crystals, this gives an
additional degree of freedom for designing a crystal which satisfied
operational parameters. A required rejection-band bandwidth can be
obtained by adjusting the magnon crystal thickness. The optimal
rejection efficiency can be obtained by adjusting the relative
groove depth.

From Fig.~\ref{diff_d}~(b) it follows that our theory is not very
accurate. The agreement is qualitative only. However, considering
the simplicity of our model, the agreement is satisfactory, and
provides key trends.

\subsection{Influence of the number of grooves}

\begin{figure}
\includegraphics[width=0.9\columnwidth]{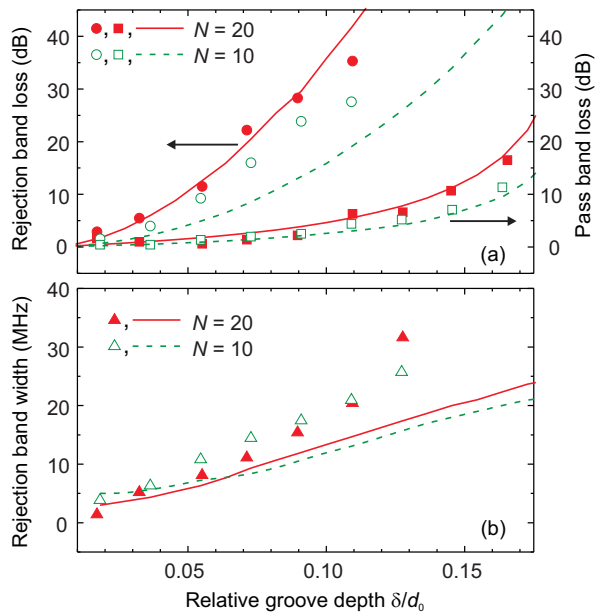}
\caption{\label{diff_N} (color online) (a) - insertion loss in the
first-order rejection (circles) and pass (squares) bands as a
function of the relative groove depth $\delta/d_0$; lines show
corresponding calculated dependencies. (b) - experimental
(triangles) and theoretical (lines) bandwidth of the first-order
rejection band as a function of relative groove depth. Filled red
symbols and solid red lines are for the magnonic crystal with $N =
20$ grooves; opened green symbols and dashed green lines are for the
crystal with $N = 10$ grooves. Film thickness $d_0 = 5.5\,\mu$m;
lattice constant $a = 300\,\mu$m; groove width $w = 30$\,$\mu$m.}
\end{figure}

The groove number is also important for the optimization of the
magnonic crystal. Fundamentally, increasing the groove number should
increase the efficiency of rejection.

According to the developed theoretical model (see Eq.~(\ref{t_mc})),
doubling the  number of grooves should double the rejection losses
on the log scale. In order to test this theoretical prediction, two
crystals with the same geometry but having different grooves numbers
$N = 10$ and $N = 20$ were investigated.

Fig.~\ref{diff_N} presents the insertion loss and the frequency
width $\Delta{f}_1$ of the rejection bands for the magnonic crystals
with the groove numbers $N = 10$  and $N = 20$, as a function of the
relative groove depth $\delta/d_0$. One sees that reducing the
number of grooves by one half results slightly decreases the
rejection efficiency. One of the possible explanation for such an
experimental behavior could be a deviation from perfect periodicity
in the lattice, which may saturate such dependence for groove
numbers larger than some characteristic value. However, a careful
examination of optical images of the quasi-crystal does not reveal
any noticeable defects. As such, we  suggest this unexpectedly weak
dependence may be connected with some peculiarity of the intrinsic
spin wave damping on the structure which is not taken into account
in our model.

Fig.~\ref{diff_N}~(b) shows experimental and calculated frequency
width of the first rejection band for the magnonic crystals with the
groove numbers $N = 10$  and $N = 20$, as a function of the relative
groove depth. No pronounced dependence on the number of groove $N$
is observed.

\subsection{Influence of the groove width}

\begin{figure}
\includegraphics[width=0.9\columnwidth]{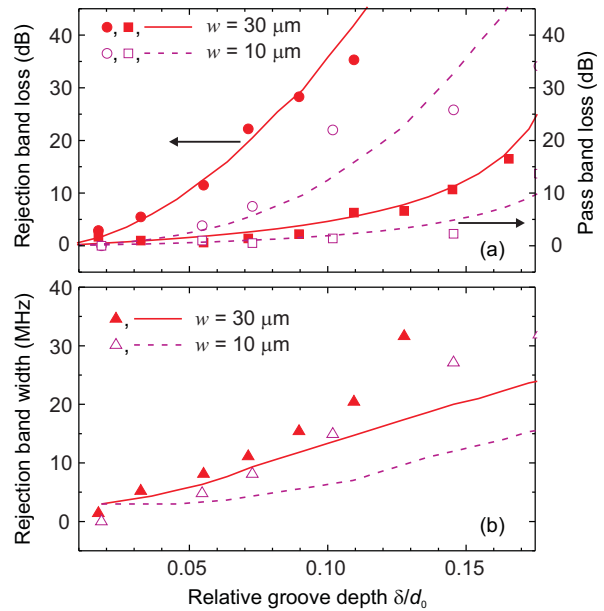}
\caption{\label{diff_w} (a) - insertion loss in the rejection band
(squares) and the pass band (circles) as a function of the groove
depth; solid lines show the respective calculated dependencies. (b)
- experimental (triangles) and theoretical (line) width of the first
rejection band as a function of the groove depth. In both panels,
filled symbols and solid lines are for a magnonic crystal with
groove width $w=30\,\mu$m; opened symbols and dashed lines are for a
magnonic crystal with groove width $w=10\,\mu$m. Film thickness $d_0
= 5.5\,\mu$m; lattice constant $a = 300\,\mu$m; number of grooves $N
= 20$.}
\end{figure}

The groove width is also an important parameter in the design and
optimization of magnonic crystals. In an ideal case, a Bragg
reflector should consist of infinitely narrow reflectors, which are
periodically placed in space. However, in our case infinitely narrow
grooves will produce no reflection. Therefore one has to keep $w$
finite. As follows from Eq.(2) and from the T-matrix theory, an
increase in the groove width increases the reflection coefficient of
individual grooves. However, this obviously introduces out-of-phase
reflections from the front and the rear edges of the grooves.
Dephasing becomes more pronounced with increasing $w$. (Indeed, such
a structure can be modeled as a superposition of Braggs reflectors
with three different lattice constants ($a, a - w,$ and $a + w$) and
lattice origins which do not coincide in space.)

To investigate the impact of the groove width, a magnonic crystal
with a groove width equal to the half of the lattice constant ($w =
a/2$) was fabricated. A suitable model for such a structure should
concept in a combination of four Bragg reflectors: two with a
lattice constant $a$, and the remaining two with lattice constant
$a/2$. Such a structure is characterized by the coincidence in the
Bragg wavenumbers between the transmission resonances for the
$a/2$-period Bragg reflector and the reflection resonances for the
$a$-period reflection. Our additional experiment investigations
proved this model: every even rejection band seen for a lattice of
narrow (30\,$\mu$m wide) grooves are absent for the structure with
$w = a/2$.

Fig.~\ref{diff_w} demonstrates the insertion loss (upper panel) and
the frequency width $\Delta{f}_1$ (bottom panel) of the first
rejection band (spin-wave wavelength here is around 600~$\mu$m) for
the magnonic crystals with groove widths $w = 10$\,$\mu$m and $w =
30$\,$\mu$m as a function of relative groove depth $\delta/d_0$. One
sees that tripling the groove width increases the rejection
efficiency by approximately 4 times in the log scale. Almost the
same effect is observed for the parasitic loss in the transmission
band (see squares in Fig.~\ref{diff_w}(a)). One can see that the
results of the calculation are in good qualitative agreement with
the experiment.

In Fig.~\ref{diff_w}(b) experimental and calculated values for the
width of the first rejection band are shown. One sees that for
different groove widths, these plots are very close to one another.
Even thought the difference is small, it may be sufficient to
fine-tune the characteristics of the magnonic crystals and enable
the obtention of required rejection efficiency and bandwidth of
rejection bands.

\section{Conclusions}

1. In this work we experimentally demonstrate that in the BVMSW
configuration a one-dimensional magnonic crystal is characterized by
the excellent spin-wave signal rejection of more than 30\,dB. The
efficiency of the rejection can be controlled by the groove depth
and width as well as the number of grooves in the crystal. A simple
theoretical model was proposed which is in good qualitative
agreement with the experimental results.

2. It is found that the optimal groove depth which ensures strong
rejection in the rejection bands while maintaining insertion loss in
the transmission bands around 3 dB is approximately 1/10 of the
total film thickness. Decreasing in the groove depth from this
optimal value leads to a drop in the rejection efficiency. With its
increasing parasitic loss in the transmission bands rapidly grows.

3. The efficiency of the rejection increases with an increasing
number of grooves. However, this increase is smaller than our
model's prediction.

4. When the width of grooves is much smaller than the spin-waves
wavelength, increasing the groove width leads to a fast increase in
the rejection efficiency.

5. The width of the rejection bands for BMVSWs exceeds the values
for the other spin-wave configurations \cite{MC review, MC skyes, MC
MSSW, MC FVMSW}. It can be controlled by the film thickness and the
groove depth. Varying the film thickness, groove depth, groove width
and number of grooves allows the engineering of magnonic crystals
with optimal characteristics.

Financial support by the DFG SE 1771/1-1, Australian Research
Council, and the University of Western Australia is acknowledged.

\end{document}